# MoS$_2$/MX$_2$ heterobilayers: Bandgap engineering via tensile strain or external electrical field

Ning Lu[1,2,⊥], Hongyan Guo[3,2,⊥], Lei Li[2], Jun Dai[2], Lu Wang[3,4], Wai-Ning Mei[4], Xiaojun Wu[3,5,*], Xiao Cheng Zeng[2,5,*]

[1]Department of Physics, Anhui Normal University, Wuhu, Anhui, 241000, China, [2]Department of Chemistry and Department Mechanics and Materials Engineering, University of Nebraska-Lincoln, Lincoln, NE 68588, USA, [3]CAS Key Lab of Materials for Energy Conversion, Department of Materials Science and Engineering, University of Science and Technology of China, Hefei, Anhui 230026, China, [4]Department of Physics, University of Nebraska-Omaha, Omaha, NE 68182, USA, [5]Hefei National Laboratory for Physical Sciences at the Microscale, University of Science and Technology of China, Hefei, Anhui 230026, China

## Abstract

We have performed a comprehensive first-principles study of the electronic and magnetic properties of two-dimensional (2D) transition-metal dichalcogenide (TMD) heterobilayers MX$_2$/MoS$_2$ (M = Mo, Cr, W, Fe, V; X = S, Se). For M = Mo, Cr, W; X=S, Se, all heterobilayers show semiconducting characteristics with an indirect bandgap with the exception of the WSe$_2$/MoS$_2$ heterobilayer which retains the direct-band-gap character of the constituent monolayer. For M = Fe, V; X = S, Se, the MX$_2$/MoS$_2$ heterobilayers exhibit metallic characters. Particular attention of this study has been focused on engineering bandgap of the TMD heterobilayer materials via application of either a tensile strain or an external electric field. We find that with increasing either the biaxial or uniaxial tensile strain, the MX$_2$/MoS$_2$ (M=Mo, Cr, W; X=S, Se) heterobilayers can undergo a semiconductor-to-metal transition. For the WSe$_2$/MoS$_2$ heterobilayer, a direct-to-indirect bandgap transition may occur beyond a critical biaxial or uniaxial strain. For M (=Fe, V) and X (=S, Se), the magnetic moments of both metal and chalcogen atoms are enhanced when the MX$_2$/MoS$_2$ heterobilayers are under a biaxial tensile strain. Moreover, the bandgap of MX$_2$/MoS$_2$ (M=Mo, Cr, W; X=S, Se) heterobilayers can be reduced by the electric field. For two heterobilayers MSe$_2$/MoS$_2$ (M=Mo, Cr), PBE calculations suggest that the indirect-to-direct bandgap transition may occur under an external electric field. The transition is attributed to the enhanced spontaneous polarization. The tunable bandgaps in general and possible indirect-direct bandgap transitions due to tensile strain or external electric field endow the TMD heterobilayer materials a viable candidate for optoelectronic applications.

[⊥]Both authors contribute equally to this work. *Electronic Emails: xzeng1@unl.edu or xjwu@ustc.edu.cn

**Introduction**

Two dimensional transition-metal dichalcogenides (TMDs) have attracted intensive interest recently owing to their novel electronic and catalytic properties that differ from their bulk counterparts.[1-3] For example, as a representative of 2D TMD materials, 2D molybdenum disulfide ($MoS_2$) monolayer possesses a direct bandgap of 1.8 - 1.9 eV while the $MoS_2$ bilayer possesses an indirect bandgap of ~1.53 eV; the $MoS_2$ transistors exhibit a high on/off ratio of $1 \times 10^8$ at room temperature. Moreover, the $MoS_2$-based integrated circuits have been fabricated and reported in the literature.[4-7]

Tunable electronic properties of 2D TMD materials are crucial for their applications in optoelectronics. Heterostructures have been widely used in conventional semiconductors for achieving tunable electronic properties. For the development of future 2D materials, the van der Waals heterostructures have been recognized as one of the most promising candidates[8] and the TMD-based hybrid multilayered structures are a prototype van der Waals heterostructures. Recently, the vertical field-effect transistor and memory cell made of TMD/graphene heterostructures have been reported.[9-12] The Moiŕe pattern of nanometer-scale $MoS_2/MoSe_2$ heterobilayer has been theoretically studied.[13] Note however that although many $MX_2$ (e.g., $MoS_2$ and $MoSe_2$) monolayers are direct-gap semiconductors, their bilayers are indirect-gap semiconductors. Recent theoretical studies suggest that the direct-bandgap character can be retained only in several heterobilayer structures[14, 15] and the heterobilayers are more desirable for optoelectronic applications.[16, 17] To achieve tunable bandgaps for 2D materials, two widely used engineering strategies are the application of either an external electric field or a tensile strain.[18-31] Previous theoretical studies have also shown that the bandgap of $MoS_2$ monolayer is insensitive to the external electric field, whereas the indirect bandgap of $MoS_2$ bilayer decreases with the increase of the external electric field.[18, 19] $MoS_2$ or $MoSe_2$ trilayer exhibits similar bandgap behavior as the bilayer counterpart when under the external electric field.[20] On the other hand, previous theoretical studies show that monolayer of TMDs can undergo the direct-to-indirect transition under the increasing tensile strain, a promising way to tune the bandgap of TMD monolayers.[21, 23] Photoluminescence spectroscopy measurements have confirmed that the optical bandgap of $MoS_2$ monolayer and bilayer decreases with the uniaxial strain and exhibits a direct-to-indirect transition.[25] Moreover, ultra high strain tenability has been demonstrated in trilayer $MoS_2$ sheets.[26] Also, under the tensile strain, the nonmagnetic $NbS_2$ and $NbSe_2$ layers can be changed to ferromagnetic.[24]

To date, most studies of TMD heterostructures are concerned about the Mo and W groups. In view of successful synthesis of nanosheets of V, Nb, Ti, Cu groups,[32-35] it is timely to examine electronic properties of TMD heterostructures and the effect of the external electric field or tensile strain on their bandgaps.[36, 37] In this study, our focus is placed on numerous $MoS_2$-based heterobilayers, including $CrS_2/MoS_2$, $CrSe_2/MoS_2$, $MoSe_2/MoS_2$, $WS_2/MoS_2$, $WSe_2/MoS_2$, $VS_2/MoS_2$, and $VSe_2/MoS_2$. For these heterobilayer systems, the lattice mismatch is typically less than 5%. We find that under an external electric field the indirect-to-direct bandgap transition may occur for two heterobilayers. A direct-to-indirect bandgap transition may occur only for the $WSe_2/MoS_2$ heterobilayer under an increasing tensile strain. In general, either the external electric field or the tensile strain can notably affect the bandgap of the TMD heterobilayers.

**Computational Methods:**

All calculations are performed within the framework of spin-polarized plane-wave density functional theory (PW-DFT), implemented in the Vienna *ab initio* simulation package (VASP).[38, 39] The generalized gradient approximation (GGA) with the Perdew-Burke-Ernzerhof (PBE) functional and projector augmented wave (PAW) potentials are used.[40-42] The effect of van der Waals interaction is accounted for with using a dispersion-corrected PBE method.[43, 44] More specifically, we adopt a $1 \times 1$ unit cell for the investigation. The vacuum size is larger than 15 Å between two adjacent images. An energy cutoff of 500 eV is adopted for the plane-wave expansion of the electronic wave function. Geometry structures are relaxed until the force on each atom is less than 0.01 eV/Å and the energy convergence criteria of $1 \times 10^{-5}$ eV are met. The 2D Brillouin zone integration using the Γ-center scheme is sampled with $9 \times 9 \times 1$ grid for geometry optimizations and $15 \times 15 \times 1$ grid for static electronic structure calculations. For each heterobilayer system, the unit cell is optimized to obtain the lattice parameters at the lowest total energy.

Biaxial tensile strain is applied to all $MX_2/MoS_2$ heterobilayers in a symmetric manner while a uniaxial tensile strain is applied in either *x*- or *y*-direction (see Figure 1 below). The direction of the external electric field is normal to the plane of heterobilayer, and in VASP, the external uniform field is treated by adding an artificial dipole sheet (i.e., dipole correction) in the supercell.[45] The geometries are kept fixed when applying the external electric field to neglect the geometric distributions to the electronic structures. The Bader's atom in molecule (AIM) method (based on charge density topological analysis) is used for charge population calculation.[46] For a few systems, the hybrid HSE06 functional is also used to confirm the trend of bandgap change.[47] In particular, the $WSe_2/MoS_2$ heterobilayer is treated as a special system for which both HSE06 calculation and PBE calculation with including the spin-orbit (SO) coupling effect[48] are reported.

**Results and Discussion**

**1. Heterobilayer of MX$_2$/MoS$_2$**

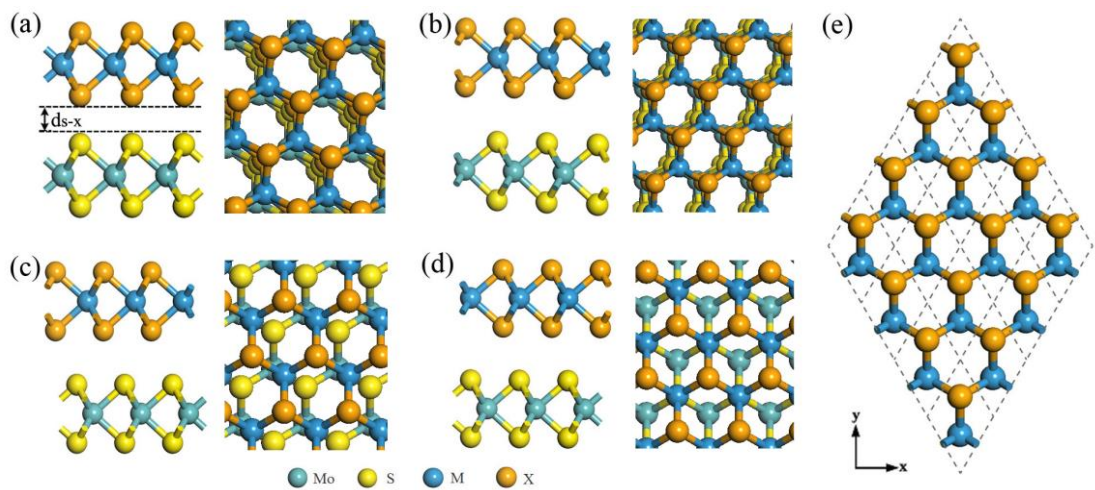

**Figure 1.** Atomic models of the MX$_2$/MoS$_2$ heterobilayer with four different types of layer-on-layer stacking: (a) AA stacking, (b) C7 stacking, (c) C27 stacking and (d) T stacking. For each stacking configuration, the left and the right panel displays the side and top view, respectively. The $d_{s-x}$ denotes the interlayer height difference between X (top-layer) and S (lower-layer) atoms. (e) The tensile strain can be applied along *x*- or/and *y*-directions.

It is known that monolayer MX$_2$ exhibits two possible structures, namely, 2H or 1T phase. The 2H structure is only considered here because it is more stable than 1T for most of the MX$_2$ structures considered in this study.[36] Moreover, following a previous study[17], we consider four different types of bilayer stacking, namely, AA, C7, C27, and T stacking, to describe how a 2H-MX$_2$ monolayer is superimposed on the 2H-MoS$_2$ monolayer (see Figure 1). A testing calculation suggests that the electronic structure is more or less the same for the four different stacking, consistent with a previous study.[17] Therefore, only the C7 stacking that gives rise to the lowest energy in most heterobilayer systems is reported for the electronic structure calculations. The optimized cell parameters and the vertical height differences between interlayer X and S atom ($d_{s-x}$, as shown in Figure 1(a)) are listed in Electronic Supplementary Information (ESI) Table S1. The $d_{s-x}$ of X and S atom in different MX$_2$/MoS$_2$ heterobilayers is less than 3.2 Å due to van der Waals interaction between MX$_2$ and MoS$_2$ layers.

The computed electronic bandgaps of $MX_2$ monolayers, bilayers, and $MX_2/MoS_2$ heterobilayers, as well as the binding energies per supercell of $MX_2/MoS_2$ heterobilayers are listed in Table 1. The binding energies are defined as $E_b$ = $E(MX_2/MoS_2$ heterobilayer$)$ − $E(MX_2$ monoayer$)$ − $E(MoS_2$ monolayer$)$, where $E(MX_2/MoS_2$ heterobilayer$)$ is the total energy of the $MX_2/MoS_2$ heterobilayer and $E(MX_2$ monoayer$)$ is the total energy of the $MX_2$ monoayer. For M = Mo, W, Cr, the $MX_2$ monolayers are direct semiconductors with the conduction band minimum (CBM) and valence band maxima (VBM) being located at the K point (ESI Figure S1). However, their corresponding bilayers become indirect semiconductors. For example, the $MoS_2$ monolayer is a direct semiconductor with a computed bandgap of 1.67 eV (PBE), while the bilayer is an indirect semiconductor with a bandgap of 1.25 eV (PBE). As shown in Figure 2, the VBM of the bilayer structures relocates to the Γ point from the K point (for the monolayer). The partial charge density at the Γ point is contributed from both monolayers, and it exhibits a strong upward shift, overtaking the energy at the K point.[14] For $MoS_2$, $WS_2$, $CrS_2$, and $CrSe_2$ bilayers, their CBM is still located at the K point. For $MoSe_2$, the CBM moves to the Λ point (Figure 2), and the energy at the Λ point is 5 meV below that at the K point. The $WSe_2$ bilayer has a nearly degenerate energy for the two valleys.

As shown in Figure 3, most $MX_2/MoS_2$ heterobilayers are indirect semiconductors, whereas only $WSe_2/MoS_2$ heterobilayer possesses a direct bandgap of 0.57 eV. Different from their own bilayers, the CBM of heterostructures are all located at the K point, while the VBM are located at Γ point. For the $WSe_2/MoS_2$ heterobilayer, however, the VBM is still located at the K point, resulting in a direct-bandgap semicondutor (PBE). The VBM of $MoSe_2/MoS_2$ at the Γ point (V1, Figure 3a) shows a mixing of densities from both monolayers. The CBM (C1, Figure 3(a)) and valence band edge (VBE, V2, Figure3a) at the K point are localized for $MoSe_2$ and $MoS_2$, respectively. The CBM and VBM positions of $MX_2$ monolayers are shown in ESI Figure S2. One can see that the band structures of $WS_2/MoS_2$, $WSe_2/MoS_2$, and $CrS_2/MoS_2$ are similar to those of $MoSe_2/MoS_2$, showing type II alignment of the band edges, which may be of advantageous for the separation of electron-hole pairs.[14]

For CrSe$_2$/MoS$_2$, the VBM at the Γ point is over that at the K point by 67 meV (Figure 3(b)), and the VBM at the Γ point is mainly due to the CrSe$_2$ layer with little contribution from the MoS$_2$ layer. However, different from other heterobilayers, the CBM and VBM at the K point are both due to the CrSe$_2$, which exhibit the type I alignment. For MX$_2$ (M=Fe, V), the monolayer, bilayer, and MX$_2$/MoS$_2$ heterobilayers all exihibit metallic character, while the ferromagnetism is still kept by the heterobilayer. As shown in Table 1, the binding energies of all the MX$_2$ and MoS$_2$ heterobilayers are in the range of -0.31 to -0.14 eV, further supporting the weak van der Waals interaction between the MX$_2$ and MoS$_2$ layers.

**Table 1**. Computed bandgap $E_{g1}$ (in eV) of the MX$_2$ monolayer, bilayer $E_{g2}$, and MX$_2$/MoS$_2$ heterobilayer $E_{g3}$, as well as the binding energies per supercell $E_b$ (in eV) of the MX$_2$/MoS$_2$ heterobilayers.

|        | $E_{g1}$ (eV) | $E_{g2}$     | $E_{g3}$      | $E_b$ (eV) |
|--------|---------------|--------------|---------------|------------|
| **MoS$_2$**  | 1.67 Direct   | 1.25 Indirect | ----          | ----       |
| **MoSe$_2$** | 1.46 Direct   | 1.20 Indirect | 0.74 Indirect | -0.16      |
| **WS$_2$**   | 1.81 Direct   | 1.43 Indirect | 1.16 Indirect | -0.22      |
| **WSe$_2$**  | 1.55 Direct   | 1.38 Indirect | 0.57 Direct   | -0.16      |
| **CrS$_2$**  | 0.93 Direct   | 0.68 Indirect | 0.39 Indirect | -0.14      |
| **CrSe$_2$** | 0.74 Direct   | 0.60 Indirect | 0.69 Indirect | -0.22      |
| **FeS$_2$**  | Metal         | metal         | metal         | -0.31      |
| **VS$_2$**   | metal         | metal         | metal         | -0.23      |
| **VSe$_2$**  | metal         | metal         | metal         | -0.16      |

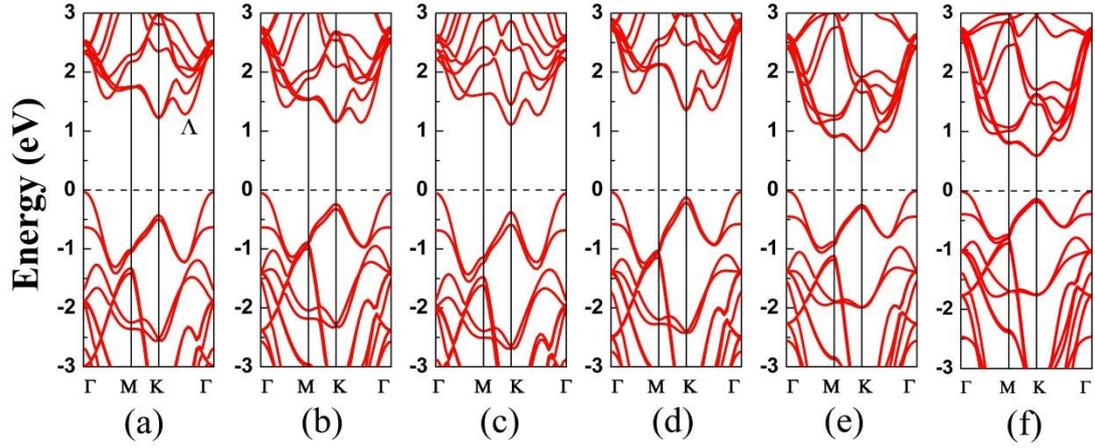

**Figure 2**. Computed band structures (PBE) of the homogeneous bilayer of (a) $MoS_2$, (b) $MoSe_2$, (c) $WS_2$, (d) $WSe_2$, (e) $CrS_2$, and (f) $CrSe_2$. All bilayers show an indirect bandgap.

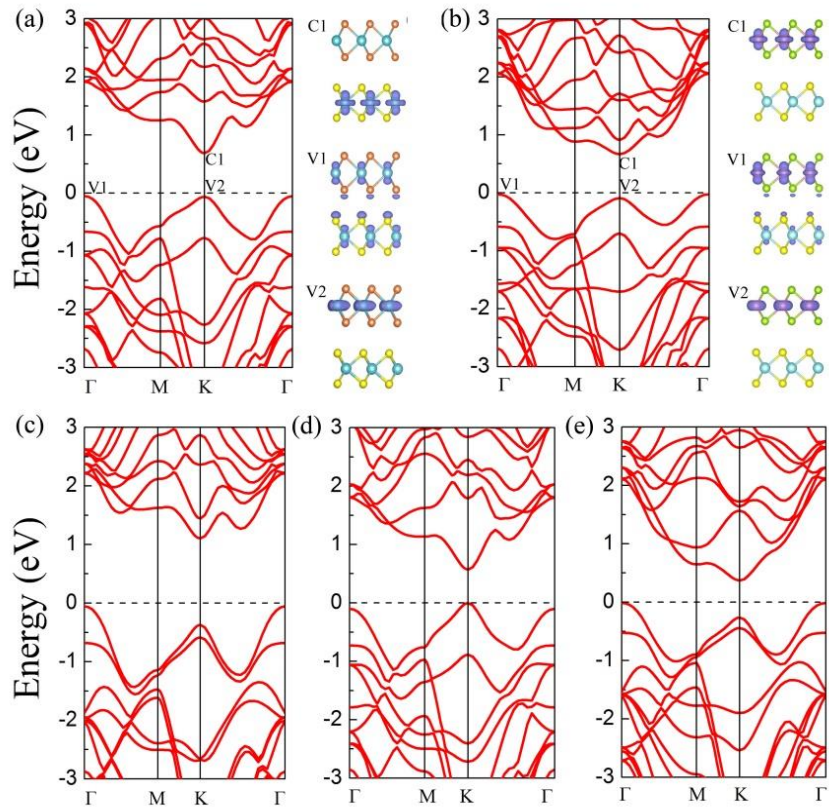

**Figure 3**. Computed band structures (PBE) and partial charge density of C1, V1, and V2 state of heterobilayer: (a) $MoSe_2/MoS_2$ and (b) $CrSe_2/MoS_2$. The isosurface value in (a) and (b) is 0.02 e/Å$^3$. Computed band structures (PES) of heterobilayer: (c) $WS_2/MoS_2$, (d) $WSe_2/MoS_2$, and (e) $CrS_2/MoS_2$. Only $WSe_2/MoS_2$ heterobilayer exhibits a *direct* bandgap.

## 2. Tunable Bandgaps via Tensile Strain

Strain modulation has been commonly used in low-dimensional systems to tune the electronic structures. For TMD monolayers, the strain-induced bandgap modification has been predicted from recent first-principles calculations.[21, 22, 24] Photoluminescence spectroscopy measurement has further confirmed the strain effect on the electronic structure of both monolayer and bilayer TMD systems. Hence, it is of both fundamental and practical interests to examine the effect of tensile strains on the electronic properties of $MX_2/MoS_2$ heterobilayers. As such, first, we have applied in-plane tensile strain by stretching the hexagonal cell biaxially,[24] and the biaxial strain is defined as $\varepsilon=\Delta a/a_0$, where $a_0$ is unstrained cell parameters and $\Delta a+a_0$ is strained cell parameters.

As mentioned above, among the heterobilayers considered in this study, only the $WSe_2/MoS_2$ heterobilayer exhibits the direct-bandgap character (Figure 3(d)). Nevertheless, we find that a 1% biaxial strain can turn the heterobilayer into an indirect semiconductor as the VBM is relocated from K to Γ point. The latter is 16 meV higher than that of the K point. The CBM is still located at the K point regardless of the strain. As the energy difference between the valence band at the K point and Γ point is just 100 meV for the unstrained $WSe_2/MoS_2$ heterobilayer, the mixing feature of the Γ point renders it more easily affected by the tensile strain. Hence, even a relatively small strain (1%) can result in higher Γ point than the K point in the energy diagram, leading to an indirect bandgap. With further increasing the biaxial strain, the energy difference between the valence band edges at these two points becomes greater. And the indirect bandgap decreases with the biaxial tensile strain, as shown in Figure 4.

The computed electronic bandgaps of the semiconducting $MX_2/MoS_2$ (M=Mo, W, Cr; X=S, Se) heterobilayers as a functional of the biaxial tensile strain is shown in Figure 4. For unstrained $MoSe_2/MoS_2$ heterobilayer, it is an indirect semiconductor with a bandgap of 0.74 eV. With the 2% biaxial tensile strain, the bandgap is reduced to 0.39 eV but still indirect. When the tensile strain increases to 4%, the bandgap is further reduced to 0.045 eV. Eventually the $MoSe_2/MoS_2$ heterobilayer turns into a metal when the biaxial strain reaches 6%. For the $WS_2/MoS_2$ heterobilayer, it turns into a metal when the biaxial tensile strain reaches 8%.

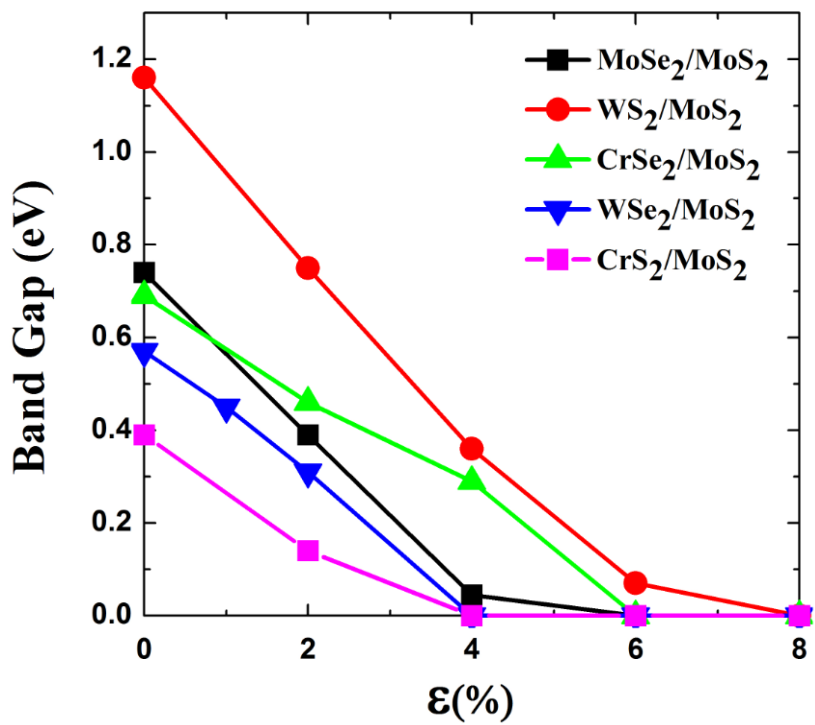

**Figure 4**. Computed electronic bandgaps (PBE) of $MX_2/MoS_2$ (M=Mo, W, Cr) heterobilayers versus the biaxial tensile strain, ranging from 0 to 8%.

As shown in Figure 4, the bandgaps of $MX_2/MoS_2$ (M=Mo, W, Cr; X=S, Se) generally decrease with the biaxial tensile strain, and undergo a semiconductor-to-metal transition at certain critical strains. To gain more insight into this transition, we have analyzed the band structures and partial density of states (PDOS) of the unstrained and strained $MX_2/MoS_2$ heterobilayer. Here, we use the PDOS of $WS_2/MoS_2$ heterobilayer as an example (see Figure 5(a)). The unstrained $WS_2/MoS_2$ heterobilayer is an indirect semiconductor with a bandgap of 1.16 eV. The VBM is mainly contributed by the $d$ orbital of W in the $WS_2$ layer, while the CBM is mainly contributed by the $d$ orbital of Mo in the $MoS_2$ layer. With a 4% biaxial tensile strain, the CBM is shifted toward the Fermi level, resulting in a reduced (indirect) bandgap (0.36 eV) for the $WS_2/MoS_2$ heterobilayer. With a 8% biaxial tensile strain, the shift of CBM leads to the semiconductor-to-metal transition (see the bottom panel of Figure 5(a)).

For the semiconducting $CrS_2/MoS_2$ heterobilayer, PBE calculation suggests that it becomes a metal with a 4% biaxial tensile strain. Here, a 2X2 supercell is used. Under a 2% biaxial tensile strain the $CrS_2/MoS_2$ heterobilayer is antiferromagnetic coupling and undergoes a nonmagnetic-to-antiferromagnetic transition. When the biaxial tensile strain increases to 10%, the $CrS_2/MoS_2$ heterobilayer turns into a strong antiferromagnetic coupling metal. Bader charge analysis suggests that the charge transfer between $CrS_2$ and $MoS_2$ layer is nearly zero under the 0% strain, and increases to 0.1e under the 10% strain, indicating that the charge transfer between $CrS_2$ and $MoS_2$ layer increases with increasing the tensile strain, leading to spontaneous polarization between $CrS_2$ and $MoS_2$ layer. In stark contrast, the $CrS_2$ monolayer cannot become magnetic even under a tensile strain as high as 15%. These results indicate that the charge transfer between $MoS_2$ and $CrS_2$ layer plays a key role in the nonmagnetic-to-antiferromagnetic transition.

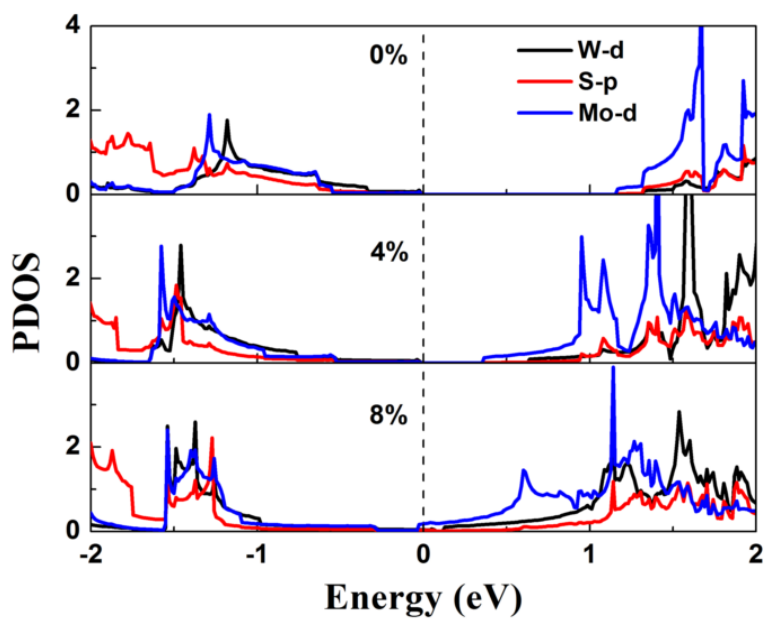

**Figure 5**. Computed partial density of states (PDOS) of the $WS_2/MoS_2$ heterobilayer under 0%, 4%, or 8% biaxial tensile strain. The vertical dashed line represents the Fermi level.

Note also that several metallic heterobilayers $MX_2/MoS_2$ (M=Fe, V; X=S, Se) still maintain their metallic character under the biaxial tensile strain. Nevertheless, we find that the magnetic moment of M and X atoms increases with the increase of the biaxial tensile strain from 0% to 10% (see Table 2). A close examination of the PDOS of $VS_2/MoS_2$ heterobilayer with 0%, 4% or 8% biaxial tensile strain (Figure 6 (a)) reveals that the state corresponding to the Fermi level is mainly contributed by *d*-states of V, which becomes more localized with increasing the strain. As shown in Figure 6(b), the spin charge density of the $VS_2/MoS_2$ heterobilayer with a 4% biaxial tensile strain suggests the magnetism is mainly contributed by the V atom (0.98 $\mu_B$) while the S atoms of $VS_2$ carry a small magnetic moment of -0.06 $\mu_B$, consistent with the analysis based on PDOS. As a result, nano-mechanical modulation of strain can turn the nonmagnetic $CrS_2/MoS_2$ heterobilayer into antiferromagnetic. The strain can also enhance the spin polarization of the $MX_2/MoS_2$ (M=Fe, V; X=S, Se) heterobilayers. This feature may be exploited in spintronic applications such as mechanical nano-switch for spin-polarized transport.

**Table 2**. Calculated magnetic moment $\mu$ ($\mu_B$) of the M and X atoms in $MX_2/MoS_2$ (M=Fe, V; X=S, Se) heterobilayers. The magnetic moment of X atoms is from $MX_2$.

| strain | $FeS_2$ | | $VS_2$ | | $VSe_2$ | |
|---|---|---|---|---|---|---|
| | Fe | S | V | S | V | Se |
| **0%** | 1.05 | -0.03 | 0.91 | -0.04 | 1.02 | -0.05 |
| **2%** | 1.50 | -0.04 | 0.94 | -0.05 | 1.05 | -0.06 |
| **4%** | 1.60 | -0.06 | 0.98 | -0.06 | 1.08 | -0.07 |
| **6%** | 1.72 | -0.07 | 1.01 | -0.07 | 1.11 | -0.08 |
| **8%** | 1.84 | -0.09 | 1.14 | -0.08 | 1.15 | -0.09 |
| **10%** | 1.98 | -0.10 | 1.19 | -0.09 | 1.18 | -0.10 |

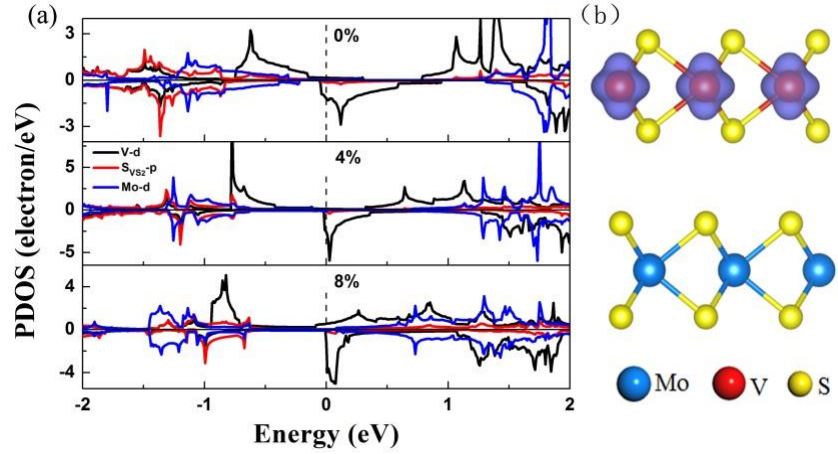

**Figure 6.** (a) Computed PDOS of $VS_2/MoS_2$ heterobilayer under 0%, 4% or 8% biaxial tensile strain. The vertical dashed line represents the Fermi level. (b) The spin charge density of $VS_2/MoS_2$ heterobilayer with a 4% biaxial tensile strain. The isosurface value is 0.01 e/Å$^3$. The blue indicates the positive values.

Besides biaxial tensile strains, we also investigate effects of a uniaxial tensile strain in either *x*- or *y*-direction (Figure 1(e)). Our calculations suggest that the bandgaps in both cases are reduced with increasing the strain, as shown in Figure 7. As mentioned above, $MoSe_2$ ($WS_2$, $CrSe_2$, $CrS_2$)/$MoS_2$ heterobilayers are indirect semiconductors. Under a uniaxial tensile strain these heterobilayers remain indirect semiconductors, the same behavior as under a biaxial tensile strain. However, the $WSe_2/MoS_2$ heterobilayer is predicted to be a direct semiconductor based on the PBE calculation. With a 2% uniaxial tensile strain along either *x*- or *y*-direction, the heterobilayer still remains to be a direct semiconductor, very different from that under the biaxial tensile strain for the heterobilayer becomes an indirect semiconductor under only 1% biaxial tensile strain. When the uniaxial tensile strain increases to 4%, the $WSe_2/MoS_2$ heterobilayer turns into an indirect semiconductor.

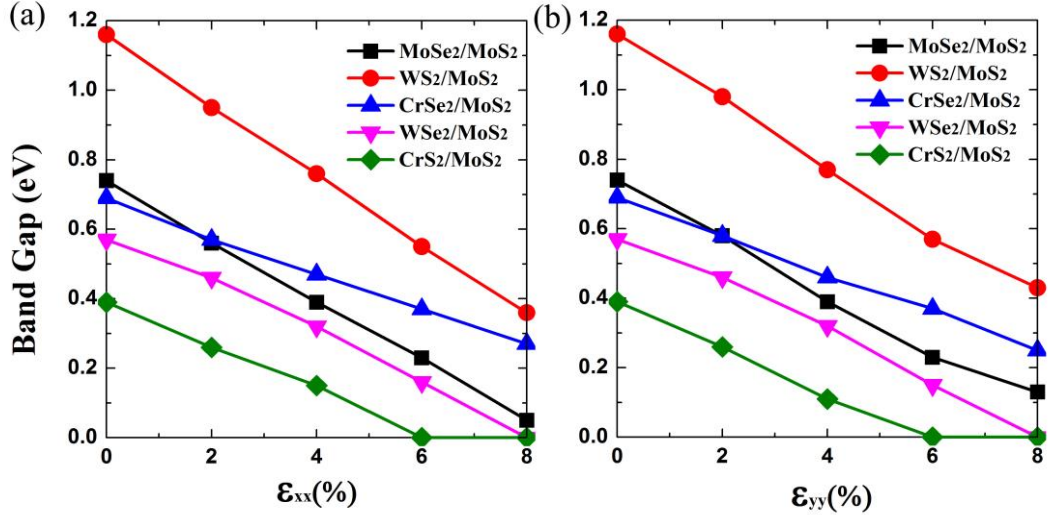

**Figure 7**. Computed bandgaps of $MX_2/MoS_2$ (M=Mo, W, Cr) heterobilayers *versus* the uniaxial tensile strain in (a) *x*- or (b) *y*-direction, ranging from 0 to 8%.

Since the $WSe_2/MoS_2$ heterobilayer is the only system here showing a direct bandgap (Figure 3(d)), additional PBE calculations with including the spin-orbit coupling effects are presented in ESI Figures S3-S5. Under either the biaxial or uniaxial tensile strain, the bandgap is still direct but much smaller. Moreover, the direct-to-indirect transition is not seen with increasing the strain. Nevertheless, the bandgap still decreases with increasing the strain and exhibits a semiconductor-to-metal transition, consistent with the PBE results. Moreover, HSE06 calculations are also performed for the $WSe_2/MoS_2$ heterobilayer. Although HSE06 tends to overestimate the bandgap (see ESI Figure S6 for a test calculation with the bilayer $MoS_2$), the overall trend in bandgap reduction with increasing the strain is the same as that predicted from the PBE calculations (see ESI Figures S3-S5). However, the direct-to-indirect transition does not occur until at the 4% biaxial strain (ESI Figure S3(l)) or 6% uniaxial strain (ESP Figures S4 and S5).

### 3. External electric field in the normal direction

$MX_2$ (M = Mo, W, Cr; X = S, Se) monolayers are direct-bandgap semiconductors, whereas their homogeneous bilayers are indirect-gap semiconductors. Importantly, among the TMD heterobilayers considered, only the WSe$_2$/MoS$_2$ heterobilayer is a direct-bandgap semiconductor, while the MoSe$_2$/MoS$_2$ heterobilayer possesses a quasi-direct bandgap with only 0.1 eV difference between the direct and indirect bandgap (Figure 3(a)), consistent with the previous study.[15] Note that the HSE06 calculation suggests that the MoSe$_2$/MoS$_2$ heterobilayer is a direct bandgap semiconductor (ESI Figure S6(b)). Previous studies also predicted direct-bandgap characters of WS$_2$/WSe$_2$ and MoTe$_2$/MoS$_2$ heterobilayers.[14, 15] We have computed the dipole moments of WSe$_2$/MoS$_2$ and MoSe$_2$/MoS$_2$ heterobilayers, and found that the dipole moments of both systems are about 0.01 e·Å greater than those of the MS$_2$/MoS$_2$ (M=Mo, W, Cr) systems, suggesting the stronger spontaneous polarization in the MSe$_2$/MoS$_2$ systems is responsible for the underlying direct-bandgap or quasi-direct-bandgap characters. This large difference in spontaneous polarization may stem from the electronegativity difference between S and Se. Assuming this explanation is valid, one could ask if an external electric field is applied to the system to increase the spontaneous polarization in MoSe$_2$/MoS$_2$, will the system undergoes an indirect-to-direct bandgap transition? Our test calculation shows that the answer to this question is yes. As shown in Figure 8(b), the applied 0.1 V/Å electric field can induce the indirect-to-direct bandgap transition in the MoSe$_2$/MoS$_2$ heterobilayer. Indeed, the VBM is moved from the Γ point to K point, and the direct transition of K-K is 0.03 eV narrower than the indirect transition of Γ-K. Further increasing the external field will reduce the direct bandgap more significantly than the indirect bandgap (Figure 8(a)). Results of a Bader charge population analysis are presented in ESI Table S2. One can see that the charge transfer between the MoS$_2$ and MoSe$_2$ layer indeed increases with the external electric field. We have also examined the bandgaps of MoSe$_2$/MoS$_2$ heterobilayer with the geometry optimized under different electric field; the results are nearly the same as those without the geometric optimization under the electric field (see ESI Table S3).

We have also examined the spontaneous polarization in the CrSe$_2$/MoS$_2$ heterobilayer which possesses a dipole moment of 0.005 e·Å. Under the external field of 0.5 V/Å, an indirect-to-direct bandgap transition is predicted. The WSe$_2$/MoS$_2$ heterobilayer always retains the direct-bandgap feature under the external electric field (Figure 8(a)), and its direct bandgap exhibits a steeper reduction with the increase of external electric field. Lastly, although the indirect-to-direct bandgap transition is not observed for WS$_2$/MoS$_2$ and CrS$_2$/MoS$_2$ heterobilayers, their indirect bandgaps exhibit a nearly linear reduction with increase of the electric field. In summary, it appears that the external electric field not only can modify bandgaps of these heterobilayers but also can induce an indirect-to-direct bandgap semiconducting transition beyond a critical field.

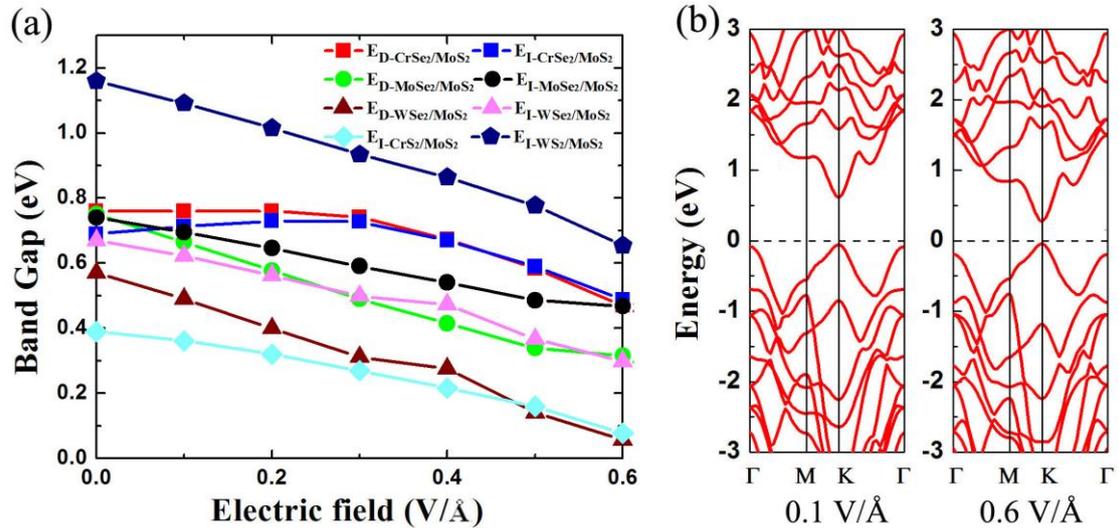

**Figure 8.** (a) Computed bandgaps (PBE) of MX$_2$/MoS$_2$ (M=Mo, W, Cr; X=S, Se) heterobilayers versus the applied electric field in the normal direction, whose strength varies from 0 to 0.6 V/ Å. $E_D$ indicates the direct bandgap of K-K transition, $E_I$ indicates the indirect bandgap of Γ-K transition. A crossover of the $E_D$ and $E_I$ curves for the heterobilayer (MoSe$_2$/MoS$_2$ and CrSe$_2$/MoS$_2$) indicates an *indirect-direct bandgap transition*. The WSe$_2$/MoS$_2$ heterobilayer is always a direct-bandgap semiconductor for field strength < 0.6 V/ Å. (b) Computed band structures of MoSe$_2$/MoS$_2$ heterobilayer under the electric field of 0.1 V/ Å or 0.6 V/ Å.

**Conclusion**

We have performed a systematic study of electronic and magnetic properties of $MX_2/MoS_2$ (M=Mo, W, Cr, Fe, V; X=S, Se) heterobilayers. Our PBE calculations suggest that $MX_2/MoS_2$ (M=Mo, W, Cr; X=S, Se) heterobilayers are indirect-bandgap semiconductors with the exception of $WSe_2/MoS_2$ heterobilayer which can retain the direct-bandgap semiconducting character. Either an external electric field or a tensile strain can induce modulation of the bandgaps for these systems. Typically, increase of the tensile strain decreases the bandgap of heterobilayers. Beyond a critical strain, the semiconductor-to-metal transition may occur. For the $WSe_2/MoS_2$ heterobilayer, a direct-to-indirect bandgap transition may occur beyond a critical biaxial or uniaxial strain; however, its bandgap is always direct regardless of the strength of external electric field (< 0.6 V/ Å). Moreover, unusual antiferromagnetism is observed in the $CrS_2/MoS_2$ system with a 2% biaxial tensile strain. The magnetic moment of M and X atoms (M=Fe, V; X=S, Se) increases with increase of the biaxial tensile strain for the $MX_2/MoS_2$ heterobilayers. The spontaneous polarization in the S/Se interface is much enhanced than the S/S interface. When an electric field is applied in the same direction as the spontaneous polarization, the indirect-to-direct bandgap semiconducting transition can be observed in two heterobilayers ($MoSe_2/MoS_2$ and $CrSe_2/CrS_2$). These theoretical predictions suggest that TMD heterobilayer materials are very promising for optoelectronic applications due to their tunable bandgaps by applying tensile strain or external electric field, possible direct-to-indirect bandgap transition in $WSe_2/MoS_2$ heterobilayer by the strain, and possible indirect-to-direct bandgap transition in $MoSe_2/MoS_2$ and $CrSe_2/CrS_2$ by the external electric field.

**Acknowledgements**

XCZ is grateful to valuable discussions with Professors Ali Adibi, Eric Vogel, Joshua Robinson, and Ali Eftekhar. The USTC group is supported by the National Basic Research Programs of China (Nos. 2011CB921400, 2012CB 922001), NSFC (Grant Nos. 21121003, 11004180, 51172223), One Hundred Person Project of CAS,

Strategic Priority Research Program of CAS (XDB01020300), Shanghai Supercomputer Center, and Hefei Supercomputer Center. UNL group is supported by ARL (Grant No. W911NF1020099), NSF (Grant No. DMR-0820521), and a grant from USTC for (1000plan) Qianren-B summer research and a grant from UNL Nebraska Center for Energy Sciences Research.

## References

1. Y. Li, Z. Zhou, S. Zhang and Z. Chen, *J. Am. Chem. Soc.,* 2008, **130**, 16739-16744.
2. K. F. Mak, C. Lee, J. Hone, J. Shan and T. F. Heinz, *Phys. Rev. Lett.,* 2010, **105**, 136805.
3. A. Splendiani, L. Sun, Y. Zhang, T. Li, J. Kim, C.-Y. Chim, G. Galli and F. Wang, *Nano Lett.,* 2010, **10**, 1271-1275.
4. B. Radisavljevic, A. Radenovic, J. Brivio, V. Giacometti and A. Kis, *Nat. Nanotechnol.,* 2011, **6**, 147-150.
5. H. Wang, L. L. Yu, Y. H. Lee, Y. M. Shi, A. Hsu, M. L. Chin, L. J. Li, M. Dubey, J. Kong and T. Palacios, *Nano Lett.,* 2012, **12**, 4674-4680.
6. B. Radisavljevic, M. B. Whitwick and A. Kis, *ACS Nano,* 2011, **5**, 9934-9938.
7. Y. Yoon, K. Ganapathi and S. Salahuddin, *Nano Lett.,* 2011, **11**, 3768-3773.
8. A. Geim and I. Grigorieva, *Nature,* 2013, **499**, 419-425.
9. S. Bertolazzi, D. Krasnozhon and A. Kis, *ACS Nano,* 2013, **7**, 3246-3252.
10. L. Britnell, R. Gorbachev, R. Jalil, B. Belle, F. Schedin, A. Mishchenko, T. Georgiou, M. Katsnelson, L. Eaves and S. Morozov, *Science,* 2012, **335**, 947-950.
11. T. Georgiou, R. Jalil, B. D. Belle, L. Britnell, R. V. Gorbachev, S. V. Morozov, Y.-J. Kim, A. Gholinia, S. J. Haigh and O. Makarovsky, *Nat. Nanotechnol.,* 2012, **8**, 100-103.
12. W. J. Yu, Z. Li, H. Zhou, Y. Chen, Y. Wang, Y. Huang and X. Duan, *Nat. Mater.,* 2012, **12**, 246-252.
13. J. Kang, J. Li, S.-S. Li, J.-B. Xia and L.-W. Wang, *Nano Lett.,* 2013, **13**, 5485-5490.
14. H.-P. Komsa and A. V. Krasheninnikov, *Phys. Rev. B,* 2013, **88**, 085318.
15. H. Terrones, F. López-Urías and M. Terrones, *Sci. Rept.,* 2013, **3**, 1549.
16. Y. Ma, Y. Dai, M. Guo, C. Niu and B. Huang, *Nanoscale,* 2011, **3**, 3883-3887.
17. K. Kośmider and J. Fernández-Rossier, *Phys. Rev. B,* 2013, **87**, 075451.
18. A. Ramasubramaniam, D. Naveh and E. Towe, *Phys. Rev. B,* 2011, **84**, 205325.
19. Q. Liu, L. Li, Y. Li, Z. Gao, Z. Chen and J. Lu, *J. Phys. Chem. C,* 2012, **116**, 21556-21562.
20. L. Kou, T. Frauenheim and C. Chen, *J. Phys. Chem. Lett.,* 2013, **4**, 1730-1746.
21. P. Johari and V. B. Shenoy, *ACS Nano,* 2012, **6**, 5449-5456.
22. Y. Ma, Y. Dai, M. Guo, C. Niu, Y. Zhu and B. Huang, *ACS Nano,* 2012, **6**, 1695-1701.
23. W. S. Yun, S. Han, S. C. Hong, I. G. Kim and J. Lee, *Phys. Rev. B,* 2012, **85**, 033305.
24. Y. Zhou, Z. Wang, P. Yang, X. Zu, L. Yang, X. Sun and F. Gao, *ACS Nano,* 2012, **6**, 9727-9736.
25. H. J. Conley, B. Wang, J. I. Ziegler, R. F. Haglund, S. T. Pantelides and K. I. Bolotin, *Nano Lett.,* 2013, **13**, 3626-3630.
26. Y. Y. Hui, X. Liu, W. Jie, N. Y. Chan, J. Hao, Y.-T. Hsu, L.-J. Li, W. Guo and S. P. Lau, *ACS Nano,* 2013, **7**, 7126-7131.
27. A. Lu, R. Zhang and S. Lee, *Appl. Phys. Lett.,* 2007, **91**, 263107.

28. Z. Zhang and W. Guo, *Phys. Rev. B,* 2008, **77**, 075403.
29. C. Zhang, A. De Sarkar and R.-Q. Zhang, *J. Phys. Chem. C,* 2011, **115**, 23682-23687.
30. Z. Zhang, W. Guo and B. I. Yakobson, *Nanoscale,* 2013, 6381-6387.
31. S. Bhattacharyya and A. K. Singh, *Phys. Rev. B,* 2012, **86**, 075454.
32. J. N. Coleman, M. Lotya, A. O'Neill, S. D. Bergin, P. J. King, U. Khan, K. Young, A. Gaucher, S. De and R. J. Smith, *Science,* 2011, **331**, 568-571.
33. J. Feng, X. Sun, C. Wu, L. Peng, C. Lin, S. Hu, J. Yang and Y. Xie, *J. Am. Chem. Soc.,* 2011, **133**, 17832-17838.
34. R. J. Smith, P. J. King, M. Lotya, C. Wirtz, U. Khan, S. De, A. O'Neill, G. S. Duesberg, J. C. Grunlan and G. Moriarty, *Adv. Maters.,* 2011, **23**, 3944-3948.
35. X. Zhang and Y. Xie, *Chem. Soc. Rev.,* 2013, **42**, 8187-8199.
36. C. Ataca, H. Sahin and S. Ciraci, *J. Phys. Chem. C,* 2012, **116**, 8983-8999.
37. H. L. Zhuang and R. G. Hennig, *J. Phys. Chem. C,* 2013, **117**, 20440-20445.
38. G. Kresse and J. Furthmüller, *Phys. Rev. B,* 1996, **54**, 11169.
39. G. Kresse and J. Furthmüller, *Comput. Mater. Sci.,* 1996, **6**, 15-50.
40. G. Kresse and J. Hafner, *Phys. Rev. B,* 1993, **47**, 558.
41. G. Kresse and D. Joubert, *Phys. Rev. B,* 1999, **59**, 1758.
42. J. P. Perdew, K. Burke and M. Ernzerhof, *Phys. Rev. Lett.,* 1996, **77**, 3865.
43. J. Klimeš, D. R. Bowler and A. Michaelides, *J. Phys.: Condens. Matter,* 2010, **22**, 022201.
44. J. Klimeš, D. R. Bowler and A. Michaelides, *Phys. Rev. B,* 2011, **83**, 195131.
45. J. Neugebauer and M. Scheffler, *Phys. Rev. B,* 1992, **46**, 16067-16080.
46. E. Sanville, S. D. Kenny, R. Smith and G. Henkelman, *J. Comput. Chem.,* 2007, **28**, 899-908.
47. J. Paier, M. Marsman, K. Hummer, G. Kresse, I. C. Gerber and J. G. Ángyán, *J. Chem. Phys.,* 2006, **125**, 249901.
48. Y.-S. Kim, K. Hummer and G. Kresse, *Phys. Rev. B,* 2009, **80**, 035203.